# A WEYL-DIRAC COSMOLOGICAL MODEL

# WITH DM AND DE

**Mark Israelit**[1]

**Abstract** In the Weyl-Dirac (W-D) framework a spatially closed cosmological model is considered. It is assumed that the space-time of the universe has a chaotic Weylian microstructure but is described on a large scale by Riemannian geometry. Locally fields of the Weyl connection vector act as creators of massive bosons having spin 1. It is suggested that these bosons, called weylons, provide most of the dark matter in the universe. At the beginning the universe is a spherically symmetric geometric entity without matter. Primary matter is created by Dirac's gauge function very close to the beginning. In the early epoch, when the temperature of the universe achieves its maximum, chaotically oriented Weyl vector fields being localized in micro-cells create weylons. In the dust dominated period Dirac's gauge function is giving rise to dark energy, the latter causing the cosmic acceleration at present. This oscillatory universe has an initial radius identical to the Plank length = 1.616 exp (-33) cm, at present the cosmic scale factor is 3.21 exp (28) cm, while its maximum value is 8.54 exp (28) cm. All forms of matter are created by geometrically based functions of the W-D theory.

## 1. Introduction

Cosmology suffers from a number of problems [1] that are in the spotlight of discussions and research. Amongst these problems are the following:

1. The Big Bang singularity of S. Weinberg's model [2] and connected difficulties, such as the flatness problem, the homogeneity and isotropy problem and the horizon (causality) problem [3]. To overcome them, complicated and theoretically insufficiently based scenarios were introduced, e.g. the inflationary scenario [4].

[1] Department of Physics and Mathematics, University of Haifa – Oranim, Tivon 36006, ISRAEL.
E-mail: israelit@macam.ac.il

2. The origin of the material content (visible matter and radiation) filling the presently observed universe is another unsolved mystery in cosmology [5]. The question is: what brought matter into being?

3. Dark matter (DM) was postulated by Fritz Zwicky [6] already in 1933 and became very popular in the last 25 years. However the nature and origin of DM is under discussion still.

4. In the end of the XX century a new phenomenon was discovered: the present cosmic acceleration [7] of our expanding universe. The factor causing this cosmic acceleration, dark energy (DE), is now widely discussed, however its nature and origin is unknown still.

Concerning the Big Bang problems, it would be interesting to remind that seventy seven years ago George Lemaître [8] introduced the term Phoenix Universe to describe an oscillatory universe (Cf. [8a]). During fifty years some models of an oscillatory universe were proposed, however the big bang model remained the popular one.

Twenty years ago in the framework of Einstein's General Theory of Relativity a singularity-free oscillating cosmological model [9] was proposed by Nathan Rosen and the present writer. In that model the spatially closed universe began its present expansion phase from a cosmic egg filled by prematter, having the huge Planckian density, $\rho_{PL} = 3.83\times 10^{65}\, cm^{-2}$, and satisfying the prematter equation of state, $P = -\rho$. The universe described in [9] achieves the maximum radius and then undergoes a contraction phase going back to the cosmic egg. The period of oscillations of that model is $\sim 1.2\times 10^{12}\, yr$. The singularity-free model [9] was further developed by F. Cooperstock and collaborators [10]. In these papers the prematter was attributed to a scalar field obeying a Klein-Gordon equation.

A "big-bounce" model with a non-vanishing cosmological constant was considered by Blome and Priester [11] and compared with the model proposed in [9]. Later a "soft bang" model of an inflationary universe without the initial singularity was proposed by Rebhan [12]. In that interesting paper various non-singular models were classified, compared and discussed.

Let us turn to the problem of matter creation. What brought matter into being? There are many works dealing with this problem from various standpoints. There are works, where the matter was created by scalar fields [10], by electromagnetic geons [13]. Very interesting are the works by Heinz Dehnen [14] and collaborators, in which microscopic Yang-Mills and Higgs fields induce matter and the gravitational field of Einstein's general theory of relativity. During the last decades became popular Wesson's Induced Matter Theory (IMT), known also as the Space Time Matter Theory, which is based on the Kaluza-Klein approach. Regarding our 4-dimensional universe as a hypersurface embedded in a 5-dimensional "bulk", Wesson and collaborators [15] have shown that the matter in our universe is induced by the geometry of the bulk, the latter being warped, but empty from matter.

Recently it was shown, that in Wesson's IMT the geometry of the 5-dimensional bulk is rather a Weylian one, than a Riemannian. The Weyl-Dirac version of Wesson's IMT was presented [16]. In that version classical models of fundamental particles induced by the bulk in our 4-dimensional universe were proposed [17]. It was also shown that conventional, dark mater and cosmic acceleration [18] may be induced in the 4-universe by the Weyl-Dirac 5-bulk.

The standpoint of the present writer is that cosmological problems have to be considered and solved in a framework that is as close as possible to Einstein's General Theory of Relativity without adding fields, which are not geometrically based. The 4-dimensional Weyl-Dirac (W-D) theory is a minimal expansion of Einstein's theory satisfying the above requirements.

In the present work we consider a classical (non quantum) 4-dimensional spatially closed cosmological model from the very beginning of the universe. In this model DM and DE are created by geometry. The work is carried out in the W-D formalism [19], [20], [21] slightly modified according to an idea suggested by Nathan Rosen [20]. The space-time is characterized by the metric tensor $g_{\mu\nu} = g_{\nu\mu}$, by the Weyl connection vector $w_\nu$ and by the Dirac gauge function $\beta$. These three geometric functions create conventional matter as well DM and DE.

We adopt the standpoint that the space-time of the universe has a chaotic Weylian microstructure (cf. [18, 23]) but is described on a large scale by the Friedmann, Lemaître, Robertson, Walker (F-L-R-W) line-element. The local field of the Weyl connection vector $w_\nu$ in micro cells will create massive bosons having spin 1. It is suggested that these bosons, called weylons, provide most of the cold dark matter (CDM) in the universe.

## 2. The Weyl-Dirac theory

The basic ideas of the Weyl-Dirac theory may be found in works of the founders [19]. Discussions and detailed descriptions as well as extensions of the W-D theory are presented in Rosen's paper [20], in the monograph by the present writer [21] and in an interesting paper by R. Carroll [21a]. Below only a concise description is given.

Weyl issued from the dominance of light rays for physical measurements. Accordingly, he regarded the light cone as the principal phenomenon describing the space-time. This idea brought Weyl to regard rather the isotropic interval $ds^2 = 0$ as invariant, than an arbitrary line-element $ds^2 = g_{\mu\nu} dy^\mu dy^\nu$ between two space-time events. In the Weyl geometry, the metric interval between two events as well the length of a given vector is no more constant, it depends on an arbitrary multiplier, the gauge function. In order to describe the geometry based on the invariance of light-cones Weyl introduced in addition to the metric tensor $g_{\mu\nu} = g_{\nu\mu}$ a length connection vector $w_\nu$. Here, both, the direction and the length of a vector change in the process of parallel displacement. If a vector having gauge-invariant contravariant components $B^\mu$ is displaced by $dx^\nu$, it changes by

$$dB^\mu = -B^\sigma \, \Gamma^\mu_{\sigma\nu} \, dx^\nu \qquad (1)$$

while the length $B = \left(B^\mu B^\nu g_{\mu\nu}\right)^{\frac{1}{2}}$ is changed by

$$dB = B \, w_\nu \, dx^\nu \qquad (2)$$

with $w_\nu$ being the Weyl connection vector. In this framework, in addition to the usual coordinate transformations appear also Weyl gauge transformations (WGT), e.g. $\tilde{g}_{\mu\nu} \to e^{2\lambda} g_{\mu\nu}$ ; $\tilde{g}^{\mu\nu} \to e^{-2\lambda} g^{\mu\nu}$ with $\lambda(x^\nu)$ being a differentiable function. For relation (2) to hold in any gauge, one must have the following WGT for $w_\mu$ :

$$w_\mu \to \tilde{w}_\mu = w_\mu + \lambda,_\mu \tag{3}$$

(A comma denotes partial differentiation)

The connection $\Gamma^\lambda_{\mu\nu}$ is expressed by the metric $g_{\mu\nu}$ and the Weyl length connection vector $w_\nu$ as

$$\Gamma^\lambda_{\mu\nu} = \left\{ {\lambda \atop \mu\nu} \right\} + g_{\mu\nu} w^\lambda - \delta^\lambda_\nu w_\mu - \delta^\lambda_\mu w_\nu \tag{4}$$

with $\left\{ {\lambda \atop \mu\nu} \right\}$ being the conventional Christoffel symbol formed from $g_{\mu\nu}$.

If a vector $B^\mu$ is transported by parallel displacement around an infinitesimal closed parallelogram one finds for the total change of the component and of the length

$$\Delta B^\lambda = B^\sigma K^\lambda_{\sigma\mu\nu} dx^\mu \delta x^\nu \quad \text{and} \quad \Delta B = B W_{\mu\nu} dx^\mu \delta x^\nu \tag{5}$$

with $K^\lambda_{\sigma\mu\nu} = -\Gamma^\lambda_{\sigma\mu,\nu} + \Gamma^\lambda_{\sigma\nu,\mu} - \Gamma^\alpha_{\sigma\mu}\Gamma^\lambda_{\alpha\nu} + \Gamma^\alpha_{\sigma\nu}\Gamma^\lambda_{\alpha\mu}$, being the curvature tensor formed of the connection $\Gamma^\lambda_{\mu\nu}$ and with the Weyl length curvature tensor given by

$$W_{\mu\nu} = w_{\mu,\nu} - w_{\nu,\mu} \tag{6}$$

EQ. (3) and (6) led Weyl [19] to identify $w_\mu$ with the potential vector, and $W_{\mu\nu}$ with the strength tensor of the electromagnetic field. In order to get a geometrically based description of gravitation and electromagnetism, Weyl derived his field equations from a variational principle $\delta I = 0$ with the action $I = \int L(-g)^{\frac{1}{2}} d^4x$. The Lagrangian density $L$ was built of the curvature tensors $K^\lambda_{\sigma\mu\nu}$, and $W_{\mu\nu}$. Weyl had to take an action, which was invariant under both coordinate transformations and WGT. For this he was forced to take an expression, which involved the square of Riemann's curvature scalar. This led to unsatisfactory equations for the gravitational field.

Dirac [19] presented a revised version of Weyl's theory. He introduced a scalar field $\beta(x^\nu)$, which under WGT changes as

$$\beta \to \tilde{\beta} = e^{-\lambda}\beta \tag{7}$$

As the scalar $\beta$ defines uniquely the gauge, it is called the Dirac gauge function. With the help of $\beta$, and making use of $K^\lambda_{\sigma\mu\nu}$ and $W_{\mu\nu}$, Dirac wrote the action integral as

$$I = \int \left[ W^{\lambda\rho} W_{\lambda\rho} - \beta^2 R + \sigma\beta^2 w^\lambda w_\lambda + 2\sigma\beta w^\lambda \beta,_\lambda + (\sigma+6)\beta,_\rho \beta,_\lambda g^{\lambda\rho} + \right. \\ \left. + 2\Lambda\beta^4 + L_{\text{matter}} \right] \sqrt{-g}\, d^4x \tag{8}$$

In (8) $\sigma$ is the Dirac parameter, $\Lambda$ is the cosmological constant and $L_{\text{matter}}$ is the Lagrangian density of matter (introduced by Rosen [20]). Varying in (8) $g_{\mu\nu}$, $w_\nu$, $\beta$ and choosing $\sigma = 0$, Dirac obtained a geometrically based theory of gravitation and electromagnetism that in the Einstein gauge, $\beta = 1$, results in Einstein's general relativity theory and Maxwell's electrodynamics.

Nathan Rosen [20], analyzing the W-D theory, showed that if one instead $\sigma = 0$ takes $\sigma < 0$ one gets a Proca [22] vector field, which from the quantum mechanical point of view may be treated as an ensemble of massive particles having spin 1. In the following we will make use of the W-D theory enriched by Rosen's approach.

### 3. The field equations

Let as turn to the action integral (8). Conventional matter is presented by the Lagrangian $L_{\text{matter}}$ that generally may depend on all three variables, $g_{\mu\nu}$; $w_\nu$ ; $\beta$, so that

$$\delta\left(L_{\text{matter}}\sqrt{-g}\right)=8\pi T^{\mu\nu}\sqrt{-g}\;\delta g_{\mu\nu}+16\pi J^{\lambda}\sqrt{-g}\;\delta w_{\lambda}+\Psi\sqrt{-g}\;\;\delta\beta\,. \qquad (9)$$

Here $T^{\mu\nu}$ is the momentum-energy tensor of conventional matter, $J^{\lambda}$ is the current of the Weyl field, $w_\lambda$ and finally, $\Psi$ stands for the charge of the Dirac gauge scalar field, $\beta$.

Varying in the action the metric tensor, $g_{\mu\nu}$ we obtain the a' la Einstein equation

$$\begin{aligned}G^{\mu\nu}=&-\frac{8\pi}{\beta^2}\left(T^{\mu\nu}+M^{\mu\nu}\right)+\frac{1}{\beta^2}\left(4\beta^{\mu}\beta^{\nu}-g^{\mu\nu}\beta^{\lambda}\beta_{\lambda}\right)+\frac{2}{\beta}\left(g^{\mu\nu}\beta^{\lambda}_{;\lambda}-\beta^{\nu}_{;\lambda}g^{\lambda\mu}\right)+\\&+\frac{\sigma}{\beta^2}\left(\beta^{\mu}\beta^{\nu}-\frac{g^{\mu\nu}}{2}\beta^{\lambda}\beta_{\lambda}\right)+\frac{\sigma}{\beta}\left(\beta^{\mu}w^{\nu}+\beta^{\nu}w^{\mu}-g^{\mu\nu}\beta_{\lambda}w^{\lambda}\right)+\\&+\sigma\left(w^{\mu}w^{\nu}-\frac{g^{\mu\nu}}{2}w^{\lambda}w_{\lambda}\right)-g^{\mu\nu}\beta^{2}\Lambda\end{aligned}\qquad (10)$$

In (10) the expression $4\pi M^{\mu\nu}=\left[\frac{1}{4}g^{\mu\nu}W^{\lambda\rho}W_{\lambda\rho}-W^{\mu}_{\ \lambda}W^{\nu\lambda}\right]$ stands for the momentum-energy tensor of the $w_\nu$ − field, while $\beta_\nu\equiv\beta_{,\nu}\equiv\frac{\partial\beta}{\partial x^{\nu}}$; $\beta^{\mu}\equiv g^{\mu\nu}\beta_{\nu}$. Taking in (10) $w_\lambda=0$; $\beta=1$ one gets the EFE of G.R.

Varying in (8) the Weyl length connection vector, $w_\mu$ yields

$$4W^{\mu\nu}_{\ ;\nu}=16\pi J^{\mu}+2\sigma\beta\left(\beta^{\mu}+\beta w^{\mu}\right) \qquad (11)$$

We assume that conventional matter has no Weylian charge

$$J^{\mu}=0 \qquad (11a)$$

so that EQ. (11) may be written as

$$W^{\mu\nu}_{\ ;\nu}=\frac{1}{2}\sigma\beta\left(\beta^{\mu}+\beta w^{\mu}\right) \qquad (11b)$$

This leads to

$$\left(\beta\beta^{\mu}+\beta^{2}w^{\mu}\right)_{;\mu}=0 \qquad (11c)$$

Finally varying in (8a) the gauge function $\beta$, one is led to the EQ.

$$R=\sigma\left(w^{\lambda}w_{\lambda}-w^{\lambda}_{;\lambda}\right)-(\sigma+6)\frac{\beta^{\lambda}_{;\lambda}}{\beta}+4\beta^{2}\Lambda+\frac{\Psi}{2\beta} \qquad (12)$$

With (11a) the contraction of EQ. (10) is identical with (12), so that Dirac's gauge function $\beta$ may be chosen arbitrarily.

According to our standpoint, the space-time of the universe has a chaotic Weylian microstructure but is on a large scale homogeneous and isotropic. In order to describe both, the microstructure and the global cosmic features, we will regard the Weyl vector as composed of two parts, $w_\nu = \underset{\text{glob}}{w_\nu} + \underset{\text{loc}}{w_\nu}$, with $\underset{\text{glob}}{w_\nu}$ being a global gradient vector, so that globally there are no vector fields. The vector $\underset{\text{loc.}}{w_\nu}$ represents the locally restricted chaotically oriented vector fields existing in micro-cells.[2]

Let us consider the global Weyl field. For the global Weyl field one has from (11b)

$$\underset{\text{glob}}{W^{\mu\nu}_{;\nu}} = \frac{1}{2}\sigma\beta\left(\beta^\mu + \beta \underset{\text{glob}}{w^\mu}\right) \tag{11d}$$

with $\underset{\text{glob}}{W_{\mu\nu}} \equiv \underset{\text{glob}}{w_{\mu,\nu}} - \underset{\text{glob}}{w_{\nu,\mu}}$ In order to get $\underset{\text{glob}}{W^{\mu\nu}_{;\nu}} = 0$ in the homogenous and isotropic universe, we take

$$\beta_\lambda + \beta \underset{\text{glob}}{w_\lambda} = 0; \quad \Rightarrow \quad \underset{\text{glob}}{w_\lambda} = -\frac{\beta_\lambda}{\beta} \tag{13}$$

(The condition $\underset{\text{glob}}{w_\nu} = 0$ would entail either $\beta = 0$ or $\sigma = 0$; both are unsuitable.)

By condition (13), EQ. (11d) is satisfied identically, while EQ. (10) takes the form

$$G^{\mu\nu} = -\frac{8\pi}{\beta^2}\underset{\text{matter}}{T^{\mu\nu}} + \frac{1}{\beta^2}\left(4\beta^\mu\beta^\nu - g^{\mu\nu}\beta^\lambda\beta_\lambda\right) + \frac{2}{\beta}\left(g^{\mu\nu}\beta^\lambda_{;\lambda} - \beta^\nu_{;\lambda}g^{\lambda\mu}\right) - g^{\mu\nu}\beta^2\Lambda \tag{14}$$

In the global equation (14), $\underset{\text{matter}}{T^{\mu\nu}}$ represents the ordinary matter as well the dark matter – weylons created in the micro cells.

## 4. Local fields, a weylon gas

Following previous work [23a, b] by Rosen and the present writer we consider the local Weyl fields. Let us turn to the field inside a microscopic cell. Due to the smallness of such a cell, the Dirac gauge function is actually constant, $\beta_\nu \equiv \frac{\partial\beta}{\partial x^\nu} = 0$, so that the Einstein gauge, $\beta = 1$, may be taken. Moreover it will be assumed that no ordinary matter is in the cell, so that from (10) we get

$$G^{\mu\nu} = -8\pi S^{\mu\nu} - g^{\mu\nu}\Lambda \tag{15}$$

$$8\pi S^{\mu\nu} = 2\left[\frac{1}{4}g^{\mu\nu}\underset{\text{loc.}}{W^{\lambda\rho}}\underset{\text{loc.}}{W_{\lambda\rho}} - \underset{\text{loc.}}{W^\mu{}_\lambda}\underset{\text{loc.}}{W^{\nu\lambda}}\right] - \sigma\left(\underset{\text{loc.}}{w^\mu}\underset{\text{loc.}}{w^\nu} - \frac{g^{\mu\nu}}{2}\underset{\text{loc.}}{w^\lambda}\underset{\text{loc.}}{w_\lambda}\right) \tag{15a}$$

[2] These chaotically oriented $\underset{\text{loc}}{w^\mu}$ -fields create massive bosons having spin 1.

Actually, $S^{\mu\nu}$ is the energy-momentum tensor of the Weyl field inside the cell. Making use of (11a) and of the condition $\beta = 1$, we rewrite EQ. (11) in the cell as

$$\underset{\text{loc.}}{W}{}^{\mu\nu}_{\ ;\nu} = \frac{1}{2}\sigma \underset{\text{loc.}}{w}{}^{\mu} \tag{16}$$

Introducing a new constant $\kappa$ given by $\sigma = -2\kappa^2$, we can rewrite with $\sigma < 0$ EQ. (16) as

$$\underset{\text{loc.}}{W}{}^{\mu\nu}_{\ ;\nu} + \kappa^2 \underset{\text{loc.}}{w}{}^{\mu} = 0 \tag{16a}$$

This is the covariant form of the Proca EQ. for a vector boson field.
From (16) follow immediately the condition

$$\underset{\text{loc.}}{w}{}^{\mu}{}_{;\mu} = 0 \tag{16b}$$,

which may be interpreted as a conservation law for the Proca current, $\kappa^2 \underset{\text{loc.}}{w}{}^{\mu}$. Inside the cell no conventional matter is present and the curvature is negligible, so that making use of $\underset{\text{loc.}}{W}_{\mu\nu} \equiv \underset{\text{loc.}}{w}_{\mu;\nu} - \underset{\text{loc.}}{w}_{\nu;\mu}$ one obtains from (16a)

$$\underset{\text{loc.}}{w}{}^{\mu}{}_{;\nu;\nu} + \kappa^2 \underset{\text{loc.}}{w}{}^{\mu} = 0 \tag{17}$$

EQ. (17) describes a vector field, that from the quantum mechanical point of view is represented by bosons of spin 1 and mass *m*, the latter in conventional units being

$$m = \frac{\kappa \hbar}{c} \tag{18}$$

It must be emphasized that the local Weyl fields existing in the cells are chaotically oriented, so that the vector sum of the spins vanishes, and only the mass effect is observed. Following [23] we assume that these bosons, named **weylons**, can make up a considerable part of cold dark matter. These weylons are analogous to photons and gravitons but differ in that they are massive. The above derived EQ-s (15), (15a), (17) and (18) are exactly the same as appear in previous works of Nathan Rosen and the present writer [23a, b]. It was shown that weylons interacting with conventionaly (luminous) matter only through gravitation may be regarded as a gas of particles in thermal equilibrium and for plausible values of mass $m$ they obey Boltzmann statistics. It was also shown that weylons having mass $m > 10\,\text{MeV}$ constitute a cold dark matter form that may be fitted into cosmological models. Supposedly these bosons were created in the very early universe, when prematter was transforming into radiation and ordinary ultrarelativistic particles. In the early universe at the time the dark matter was created, it had a density that was negligibly small compared to that of ordinary matter. The fractional abundance of weylon dark matter became important at the time when galaxies were formed, and it may be that weylon dark matter, through its gravitational interaction with the ordinary matter played an important, stimulating role in this process.

We will discuss the emergence and behavior of weylon DM considering the radiation and dust dominated periods of the global field (SEC-s. 5e, 7)

## 5. The global field

As mentioned above for cosmological considerations in the homogeneous, isotropic universe we take condition (13), so that EQ. (11d) is satisfied identically, and we are left with the gravitational EQ. (14). We recall that $\underset{\text{matter}}{T^{\mu\nu}}$ represents conventional matter as well dark matter – weylons created in the micro cells.

Let us consider a closed homogeneous and isotropic universe described by the F-L-R-W line element

$$ds^2 = dt^2 - a^2(t)\left[\frac{dr^2}{1-r^2} + r^2 d\vartheta^2 + r^2 \sin^2 \vartheta d\varphi^2\right] \qquad (19)$$

with $a(t)$ being the cosmic scale parameter. By the assumed symmetry of the universe, Dirac's gauge function depends only on the cosmic time $t$.
Taking into account (19) and inserting into (14) $\beta(t)$ we obtain the EFE explicitly

$$G_0^0 = -8\pi \underset{\text{glob}}{\rho} = -\frac{8\pi}{\beta^2}\underset{\text{matter}}{\rho} + 3\frac{(\dot{\beta})^2}{\beta^2} + 6\frac{\dot{a}}{a}\frac{\dot{\beta}}{\beta} - \beta^2\Lambda \qquad (20a)$$

$$G_1^1 = 8\pi \underset{\text{glob}}{P} = \frac{8\pi}{\beta^2}\underset{\text{matter}}{P} - \frac{(\dot{\beta})^2}{\beta^2} + 2\frac{\ddot{\beta}}{\beta} + 4\frac{\dot{a}}{a}\frac{\dot{\beta}}{\beta} - \beta^2\Lambda \qquad (20b)$$

Here a dot denotes partial derivative with respect to $t$ and $\underset{\text{matter}}{\rho}$ and $\underset{\text{matter}}{P}$ stand for the density and pressure of both, conventional matter and weylon DM. From (20a, b) we obtain

$$\frac{\dot{a}^2}{a^2} = \frac{8\pi}{3\beta^2}\underset{\text{matter}}{\rho} - \frac{(\dot{\beta})^2}{\beta^2} - 2\frac{\dot{a}}{a}\frac{\dot{\beta}}{\beta} + \frac{1}{3}\beta^2\Lambda - \frac{1}{a^2} \qquad (21a)$$

$$\frac{\ddot{a}}{a} = -\frac{4\pi}{\beta^2}\left(\underset{\text{matter}}{P} + \frac{1}{3}\underset{\text{matter}}{\rho}\right) + \frac{(\dot{\beta})^2}{\beta^2} - \frac{\dot{a}\dot{\beta}}{a\beta} - \frac{\ddot{\beta}}{\beta} + \frac{1}{3}\beta^2\Lambda \qquad (21b)$$

From the Bianchi identity one has

$$\underset{\text{glob}}{\dot{\rho}} = -3\frac{\dot{a}}{a}\left(\underset{\text{glob}}{\rho} + \underset{\text{glob}}{P}\right) \qquad (22)$$

Making use of (20a, b) one obtains the energy relation

$$\frac{8\pi}{\beta^2}\underset{\text{matter}}{\dot{\rho}} = -\frac{3\dot{a}}{a}\frac{8\pi}{\beta^2}\left(\underset{\text{matter}}{\rho} + \underset{\text{matter}}{P}\right) + \frac{2\dot{\beta}}{\beta^3}8\pi\underset{\text{matter}}{\rho} + 6\left(\frac{\dot{\beta}\ddot{\beta}}{\beta^2} - \frac{(\dot{\beta})^3}{\beta^3} + \frac{\ddot{a}\dot{\beta}}{a\beta} + \frac{\dot{a}(\dot{\beta})^2}{a\beta^2}\right) - 2\dot{\beta}\beta\Lambda \qquad (23)$$

In equations (20-21) Dirac's gauge function $\beta(t)$ is involved. This can be chosen nearly arbitrarily in the W-D framework; the only restriction is $\beta(t) > 0$.

### 5-a. Choosing a certain gauge function

It is convenient to introduce the Dirac gauge function $\beta_{\text{total}}$ as formed of two parts,

$$\beta_{\text{total}} = U_1\,\beta_{\text{create}} + U_2\,\beta_{\text{late}} \qquad (24)$$

where $U_1$ and $U_2$ are step functions given by:

$$U_1 = \left\langle \begin{array}{l} 1, \; for \; a_0 < a(t) < A \\ \frac{1}{2}, \; for \; a = A \\ 0 \; for \; a(t) > A \end{array} \right. \quad \text{and} \quad U_2 = \left\langle \begin{array}{l} 0, \; for \; a(t) < A \\ \frac{1}{2}, \; for \; a = A \\ 1, \; for \; a(t) > A \end{array} \right. \qquad (24a)$$

Making use of present cosmological observations we can fix $A = 1.544 \times 10^{-3}\text{cm}$ (Cf. below Sec.5d, 6, 7). In EQ. (24a), $a(t)$ is the cosmic scale parameter and $a_0$ is that of the universe at the very beginning. Further, $\beta_{\text{create}}$ is the gauge function in the very early universe, while $\beta_{\text{late}}$ stands for that function at $a(t) > \text{A}$.

Let us choose $\beta_{\text{create}}(t)$ according to the following scenario. We suppose that the universe began its present expansion phase from a sphere having radius $a_0$ identical with the Planckian length, i.e. $a_0 \equiv l_P = 1.616 \times 10^{-33}\text{cm}$; this initial state was stationary and at the very beginning no matter (ordinary or weylon DM) was present. There was only the Dirac gauge function $\beta_{\text{create}}(t)$. We assume that $\beta_{\text{create}}(t)$ had a huge maximum at the beginning ($t = 0$), that it is an even function of time and that it is rapidly decreasing. In the very early universe, $\beta_{\text{create}}$ is a matter creating function. There are of course many functions satisfying the above requirements. We will adopt the following:

$$\beta_{\text{create}} = \frac{\text{B}}{\cosh^n(\gamma t)} \qquad (25)$$

In (25), B (Capital Beta) and $\gamma$ are positive constants and $n$ is a positive integer. These constants will be fixed below.
In the Appendix are given the derivatives of $\beta_{\text{create}}$ (Cf. (A-1) – (A-3)).

The function $\beta_{\text{late}}$ acts from the moment, when the radius of the universe $a(t) > A$. Its influence is almost negligible up to the dust dominated period, where it creates dark energy, the latter being responsible for the acceleration of our universe at present.

Let us take the gauge function $\beta_{\text{late}}$ as (Cf. [26])

$$\beta_{\text{late}} = 1 + C_1 \tanh[\delta_1 (X - 1)] + C_2 \tanh\left[\delta_2 \left(X^2 - 1\right)\right] \qquad (26)$$

Here $X(t) = \dfrac{a(t)}{a_N}$, and $a_N$ stands for the present value of the cosmic scale parameter.

In Appendix are given the derivatives of $\beta_{\text{late}}$ (Cf. (B-1), (B-2)). The parameters $C_1; C_2; \delta_1; \delta_2$ will be fixed below in accordance with cosmological observations.

## 5-b. The beginning

Let us consider the situation at the very beginning, $t = 0$, when $\rho_{\text{matter}} = 0$; $P_{\text{matter}} = 0$, (no matter present) . Making use of (25) and of (A-1) – (A-4) we rewrite (21a) as

$$\left(\frac{\dot{a}^2}{a^2}\right) = \frac{1}{3}\text{B}^2\Lambda - \frac{1}{a_0^{\;2}} \qquad (27)$$

Supposing that at $t=0$ there was a stationary state, $\left(\frac{\dot{a}}{a}\right)_{/t=0}=0$, we obtain

$$\frac{1}{3}\mathrm{B}^2\Lambda=\frac{1}{a_0^{\ 2}} \tag{27a}$$

Here $a_0\approx l_\mathrm{P}=1.616\times10^{-33}\mathrm{cm}$ and according to **[25]** $\Lambda\approx2.074\times10^{-58}\mathrm{cm}^{-2}$**.** so that

$$\mathrm{B}^2\approx0.554\times10^{124};\ \mathrm{B}\equiv\beta_{\mathrm{create}}(t=0)\approx0.744\times10^{62} \tag{28}$$

Going back to EQ. (21a) we see that close to the beginning it may be written as $\dot{a}^2=\frac{1}{3}\mathrm{B}^2\Lambda\,a^2-1$, which is satisfied by

$$a(t)=a_0\cosh\left(\frac{t}{a_0}\right) \tag{29}$$

Let us turn to EQ. (21b). For the very early universe it takes the form

$$\frac{\ddot{a}}{a}=-\frac{\ddot{\beta}}{\beta}+\frac{1}{3}\beta^2\Lambda \tag{30}$$

Making use of (A-1) – (A-4) we are led by (27a) and (30) to the condition

$$-\left(\frac{\ddot{\beta}}{\beta}\right)_0\doteq n\gamma^2<<\frac{1}{3}\mathrm{B}^2\Lambda=\frac{1}{a_0^2}=3.83\times10^{65}\ cm^{-2} \tag{30a}$$

At the very beginning in the universe acts a huge acceleration

$$\frac{\ddot{a}}{a}\approx\frac{1}{3}\mathrm{B}^2\Lambda=\frac{1}{a_0^2}=3.83\times10^{65}\ cm^{-2} \tag{30b}$$

This acceleration will cause a high speed expansion.

## 5-c. Creation of matter, the prematter period

It can be shown that very close to the beginning the Dirac gauge function $\beta_{\mathrm{create}}$ will create matter. Indeed, taking into account $\rho_{\mathrm{matter}}=0;\ P_{\mathrm{matter}}=0$ one has from (23)

$$\frac{8\pi}{\beta^2}\dot{\rho}_{\mathrm{matter}}=6\left(\frac{\dot{\beta}\ddot{\beta}}{\beta^2}-\frac{(\dot{\beta})^3}{\beta^3}+\frac{\ddot{a}\dot{\beta}}{a\beta}+\frac{\dot{a}(\dot{\beta})^2}{a\beta^2}\right)-2\dot{\beta}\beta\Lambda \tag{31a}$$

Expanding $\beta_{\mathrm{create}}(t);\ \frac{\dot{a}}{a};\ \frac{\ddot{a}}{a}$ in time, one obtains from (31a) for $\gamma t<<1$ the following matter creation relation

$$\dot{\rho}_{\mathrm{matter}}=\frac{3}{4\pi}\mathrm{B}^2n^2\gamma^4t; \tag{32}$$

Thus, matter created during a very small time, $t$, after the beginning has the density

$$\rho_{\mathrm{matter}}(t)=\frac{3}{8\pi}\mathrm{B}^2n^2\gamma^4t^2 \tag{32a}$$

We adopt the idea that there exists a limit value for $\underset{\text{matter}}{\rho}$, the Planck density[3] $\rho_{\text{Pl}} = 3.83\times 10^{65}\text{cm}^{-2}$ (that is equal to $\frac{1}{3}\text{B}^2\Lambda$), so that, after achieving this limit density, the growing of matter density will be stopped: $\underset{\text{matter}}{\rho}$ has a saturation point. The moment, when that saturation takes place is unknown. Let as assume that it occurs very close to the beginning, at $t_{\text{Pl}} = 5a_0 = 8.08\times 10^{-33}\text{cm}$, when according to (29) the radius is $a_{\text{Pl}} = 1.2\times 10^{-31}\text{cm}$. Actually, $t_{\text{Pl}}$ and $a_{\text{Pl}}$ describe the beginning of the matter universe with $\rho_{\text{Pl}}$. With $t_{\text{Pl}}$ and $\rho_{\text{Pl}}$ one obtains from (32a) $n\gamma^2 \approx 2.98\times 10^3\text{cm}^{-2}$ a value that satisfies condition (30a). The value of the power $n$ is arbitrary. It turns out that $\underline{n = 1000}$ is a good choice, so that $\underline{\gamma = 1.726\text{cm}^{-1}}$. The underlined here quantities will be hereafter adopted.

After $t_{Pl}$, when the sphere is filled with matter, EQ. (31a) is no more holding. With nonzero $\underset{\text{matter}}{P}$ ; $\underset{\text{matter}}{\rho}$ one obtains from (23) and (21a,b) the following energy relation

$$\underset{\text{matter}}{\dot{\rho}} = -3\left(\frac{\dot{a}}{a} + \frac{\dot{\beta}}{\beta}\right)\left(\underset{\text{matter}}{\rho} + \underset{\text{matter}}{P}\right) + 4\frac{\dot{\beta}}{\beta}\underset{\text{matter}}{\rho} \tag{31b}$$

As we know nothing about the relation between pressure and density we turn to EQ-s (21a,b). Making use of (29) as well (A-1) – (A-3) we obtain in the very early universe the following equation of state (EoS)

$$\underset{\text{matter}}{P} = -\underset{\text{matter}}{\rho} \tag{33}$$

But this is exactly the EoS for prematter, a substance introduced in [9]. We assume that the prematter period lasts up to $t_1 = 65a_0 = 1.05\times 10^{-31}\text{cm}$.

By (33) we have from (31b)

$$\underset{\text{matter}}{\dot{\rho}} = 4\frac{\dot{\beta}_{create}}{\beta_{create}}\underset{\text{matter}}{\rho} \tag{34}$$

Making use of (A-3) we obtain $\underset{matter}{\rho} = \rho_{Pl}\cdot e^{-4n\gamma^2 t}$, from which it follows that during the whole prematter period, $t_{Pl} = 5a_0 \le \underset{\text{prematter}}{t} \le t_1 = 65a_0$, the matter density remains equal to $\rho_{\text{Pl}}$. One can prove that EQ. (29) is holding during the whole prematter period, so that at its end, when $t_1 = 1.05\times 10^{-31}\,cm$, one has $a_1 = a(t_1) = 1.37\times 10^{-5}\,cm$.

### 5-d. The transition period

The prematter period is followed by a period, during which the prematter turns gradually into radiation. This transition period lasts from $t_1 = 1.05\times 10^{-31}\,cm$;

---

[3] We take $\rho_{\text{Pl}} = 3.83\times 10^{65}\text{cm}^{-2}$ in geometric quantities. In conventional quantities one has $\rho_{\text{Pl}} = 5.16\times 10^{93}\frac{\text{g}}{\text{cm}^3}$

$a_1 = 1.37\times 10^{-5} cm$, with $\beta_{\text{create}} = \mathrm{B}$ and $\rho_{\text{matter}} = \rho_{\text{Pl}}$, up to $a_{\text{Rad}} = 100\text{cm}$, when the universe is filled by radiation and ultrarelativistic particles. Hereafter we will drop the subscript $_{\text{matter}}$ that appeared in $\rho_{\text{matter}}$ and $P_{\text{matter}}$.

During the transition period, $\beta_{\text{create}}(t)$ and $\rho$ are decreasing gradually; accordingly the EoS is changing from the prematter one $P = -\rho$ to the radiation $P = \frac{1}{3}\rho$. Let us write for the transition period the following EoS (Cf. [9])

$$P = \frac{1}{3}\rho\left(1 - 4\frac{\rho}{\rho_{\text{Pl}}}\right) \tag{35}$$

so that if $\rho$ is close to $\rho_{\text{Pl}}$, EQ. (35) turns into (33), while for $\rho << \rho_{\text{Pl}}$ it gives the radiation EoS

$$P = \frac{1}{3}\rho \tag{36}$$

Let as go to the energy relation (31b). Making use of (35) and taking into account that during the whole transition period one has $\left|\frac{\dot{\beta}}{\beta}\right| << \frac{\dot{a}}{a}$, we can rewrite (31b) as

$$\dot{\rho} + 4\frac{\dot{\text{a}}}{\text{a}}\left(\rho - \frac{\rho^2}{\rho_{Pl}}\right) = 0 \tag{31c}$$

Integrating one obtains for the matter density

$$\rho = \frac{A^4\rho_{Pl}}{\mathrm{A}^4 + a^4} \tag{37}$$

Further, making use of (35) one has for the pressure

$$P = \frac{A^4\rho_{Pl}\left(\frac{1}{3}\cdot a^4 - A^4\right)}{\left(A^4 + a^4\right)^2} \tag{37a}$$

In (37, 37a) the constant of integration $A$ will be chosen as the radius of the universe at a moment inside the transition period, so that for $a << A$ equation (33) is valid, while for $a >> A$ (36) holds. The value of $A$ may be linked to present cosmological observations. It is believed (Cf. [27]) that the present overall, $\rho_{\text{tot}} = 7.3488\times 10^{-58}\text{cm}^{-2}$, consists of $5\%$ ordinary; $23\%$ CDM; $72\%$ DE$(\beta + \Lambda)$ matter. From this one can estimate $A = 1.544\times 10^{-3}\text{cm}$. Making use of (29) one has for the according moment $t_A = 69.7a_0 \approx 1.127\times 10^{-31}\text{cm}$. At this state there is still $\rho = \rho_{\text{Pl}}$.

In order to find $a(t)$ at any $t$, one can go back to EQ. (21a). Inserting (37) one obtains

$$\frac{\dot{a}}{a} + \frac{\dot{\beta}}{\beta} = \sqrt{\frac{8\pi}{3\beta^2}\cdot\frac{A^4\rho_{\text{Pl}}}{\mathrm{A}^4 + a^4} + \frac{1}{3}\beta^2\Lambda - \frac{1}{a^2}} \tag{38}$$

In the transition period $\left|\frac{\dot{\beta}}{\beta}\right| << \frac{\dot{a}}{a}$, so that $\frac{\dot{a}}{a} = \sqrt{\frac{8\pi}{3\beta^2}\cdot\frac{A^4\rho_{\text{Pl}}}{\mathrm{A}^4 + a^4} + \frac{1}{3}\beta^2\Lambda - \frac{1}{a^2}}$

Considering the times, when $\beta \approx \mathrm{B}$, one has $\frac{\dot{a}}{a} = \sqrt{\frac{1}{a_0^2} - \frac{1}{a^2}}$ , so that

$$a(t) = a_0 \cosh\left(\frac{t}{a_0} + Const\right). \tag{38a}$$

The *Const* in (38a) may be used to sew the early transition period, when $\beta \approx \mathrm{B}$, with the later one, when $\beta < 10$

For $a > A = 1.544 \times 10^{-3}\mathrm{cm}$ when $\beta_{\mathrm{create}} = 0$; $\beta_{\mathrm{late}} \approx 1$ (Cf. Appendix) one can rewrite (38) as

$$\frac{\dot{a}}{a} = \sqrt{\kappa^2 \frac{A^4}{\left(a^4 + A^4\right)} - \frac{1}{a^2}} \tag{38b}$$

with $\kappa^2 = \frac{8\pi}{3}\rho_{Pl} = 3.2 \times 10^{66}\mathrm{cm}^{-2}$. This may be rewritten as $\frac{\dot{a}}{a} \approx \kappa \sqrt{\frac{a^2}{\left(\frac{a^4}{A^4} + 1\right)a^2}}$

Introducing $x \equiv \frac{a}{A}$, one obtains the following relation for $x(t)$

$$\left.\frac{1}{2}\sqrt{1 + x^4} - \frac{1}{2}\ln\frac{\sqrt{1 + x^4} + 1}{x^2}\right|_1^{\frac{a}{A}} = \kappa(t - t_A) \tag{39}$$

With $A = 1.544 \times 10^{-3}\mathrm{cm}$ and $t_A = 1.127 \times 10^{-31}\mathrm{cm}$ one can obtain $a(t)$. So one has $a = 2A$ at $t = t_A + 1.38 \times 10^{-33} \approx 1.44 \times 10^{-31}\mathrm{cm}$. An interesting state is at $a^4 = 3A^4$, where according to (37a) $P = 0$: here the pressure is passing from negative values (rather tension, than pressure) to positive ones. At $t \approx 3.05 \times 10^{-30}\mathrm{cm}$ and $a = 100A = 1.544 \times 10^{-1}\mathrm{cm}$ one has already the radiation EoS (Cf. (37) and (37a)). The end of the transition, beginning of the radiation dominated period we take when $a_{\mathrm{Rad}} = 100\mathrm{cm}$; and according to (39) the time is $t_{\mathrm{Rad}} = 1.172 \times 10^{-24}\mathrm{cm}$.

**5-e. Temperature in the early universe. Emergence of weylon DM**

The state $t_{\mathrm{Rad}}$ ; $a_{\mathrm{Rad}}$ with the EoS $P = \frac{1}{3}\rho$, will be taken as the beginning of the radiation dominated period, which will last up to a state $a_{\mathrm{Dust}}$ with most of the matter already being in the state of dust. Below (Sec. 6.) it will be shown that the present value of the cosmic scale parameter $a_N = 3.208 \times 10^{28}\mathrm{cm}$. It seems that as a plausible value for the beginning of the dust dominated era may be taken $a_{\mathrm{Dust}} = 10^{-2} a_N \approx 3.208 \times 10^{26}\mathrm{cm}$; this will be adopted hereafter.

Let us consider the dependence of temperature *T* on the scale parameter $a$ in the early universe. Following [1] and [9] we regard density and pressure as functions of temperature, i. e. $\rho = \rho(T)$, $P = P(T)$ and write for the entropy

$$dS(V,T) = \frac{1}{T}\left(d(\rho V) + P\,dV\right) \tag{40}$$

with $V = 2\pi^2 a^3$ being the volume of our closed universe. From EQ. (40) follows

$$\frac{\partial S}{\partial V} = \frac{1}{T}(\rho + P) \quad \text{and} \quad \frac{\partial S}{\partial T} = \frac{V}{T}\frac{\partial \rho}{\partial T} \tag{41}$$

with the integrability condition

$$\frac{dP}{dT} = \frac{1}{T}(\rho + P) \tag{42}$$

Inserting the EoS (35) into (42) and integrating we obtain

$$\rho\left(1 - \frac{\rho}{\rho_{\mathrm{Pl}}}\right)^7 = \sigma T^4 \tag{43}$$

where the constant of integration is taken equal to the Stefan-Boltzmann constant, $\sigma = 6.24 \times 10^{-64}\mathrm{cm}^{-2}(\mathrm{K})^{-4}$. Making use of the expression (37) for $\rho$ we obtain

$$T = \left(\frac{\rho_{\mathrm{Pl}}}{\sigma}\right)^{1/4} \frac{A a^7}{\left(A^4 + a^4\right)^2} \tag{44}$$

With $\left(\frac{\rho_{\mathrm{Pl}}}{\sigma}\right)^{1/4} = 1.574 \times 10^{32}\mathrm{K}$ and $A = 1.544 \times 10^{-3}\mathrm{cm}$, one can account the temperature in the early universe at various values of the cosmic scale parameter.

At the beginning of the matter universe ($a_{\mathrm{Pl}} = 1.2 \times 10^{-31}\mathrm{cm}$) the universe was extremely cold $T_{\mathrm{Pl}} = 2.70 \times 10^{-165}\mathrm{K}$, at the end of the prematter period ($a_1 = 1.37 \times 10^{-5}\mathrm{cm}$) one had $T_1 = 5.24 \times 10^{17}\mathrm{K}$. The temperature continues to increase in the transition period and when $a^4 = 7A^4$ we get $T_{\max} = 7.41 \times 10^{31}\mathrm{K}$. From this point on $T$ decreased. At the beginning of the radiation period ($a_{\mathrm{Rad}} = 100\,\mathrm{cm}$) there was $T_{\mathrm{Rad}} = 2.43 \times 10^{27}\mathrm{K}$.

Let us go back to Sec. 4 and consider the emergence and development of weylon DM. Following [23a], [23b] we assume that weylons were created in the Weylian micro-cells while prematter was transforming into radiation and ordinary ultrarelativistic particles, and the temperature was close to its maximum value $T_{\max} = 7.4 \times 10^{31}\mathrm{K}$. It is plausible to assume that the creation process ended at $T_{\mathrm{Rad}} = 2.43 \times 10^{27}\mathrm{K}$, when prematter was completely transformed into radiation. At this moment both, the weylons and conventionally matter had the same temperature, $T_{\mathrm{Rad}}$. Thereafter, the weylons did not interact with conventionally (luminous) matter, they developed in time independently of the latter and their total number in the universe remained constant, i. e. $n_w a^3 = N_{\mathrm{w.total}} = \mathrm{const}$. During the radiation period, the temperature of conventionally matter, $T_{\mathrm{conv.}} \propto a^{-1}$, and at the beginning of the dust period $a_{\mathrm{Dust}} \approx 3.208 \times 10^{26}\mathrm{cm}$ the temperature was $T_{\mathrm{Dust.conv.}} = 7.575 \times 10^2 K$. On the other hand the weylons constitute a gas of heavy particles, $10^5 GeV \geq m \geq 10\,\mathrm{MeV}$ (Cf. [23]) and their temperature $T_W$ decreases rapidly. At present the conventionally matter is at $T_{N.conv.} \approx (2.7 - 3.1)K$, whereas the temperature of the weylon gas, that depends on the weylon mass, is in the interval $10^{-17} K \leq (T_W)_N \leq 10^{-10} K$ . We will adopt the results of [23], where it was shown that weylons constitute a cold dark matter form that may be fitted into cosmological models. In the early universe, at the time the weylon DM was created, it had a density that was negligibly small compared to that of ordinary matter. The fractional abundance of weylon dark matter became important at the time when galaxies were formed; at present $\rho_{weylon\,DM} / \rho_{convent.m.} \approx 4.61$.

## 6. The present state. Specifying the Dirac gauge function $\beta_{\text{late}}$

In EQ. (26) we have introduced the Dirac gauge function $\beta_{\text{late}}$ that acts from $a(t) > A = 1.544 \times 10^{-3}\,\text{cm}$ up to $a_{\max}$. In the dust dominated period this function will create DE, the latter causing the present cosmic acceleration. It is important to note that in the early universe up to the beginning of the dust dominated era $\beta_{\text{late}} \approx 1$, while its derivatives take insignificant values (Cf. (B-10), (B-11)). The function $\beta_{\text{late}}$ depends on four parameters $C_1; C_2; \delta_1; \delta_2$, as well on the present value of the scale parameter $a_N$. Below these parameters will be fixed by present cosmological observations as well by the requirement that in the early universe $\beta_{\text{late}}(X << 1) \approx 1$, the latter leading to the condition

$$C_1 \tanh[\delta_1] + C_2 \tanh[\delta_2] = 0 \qquad (45)$$

From (26) one has at present $\beta_{\text{late}}(X = 1) = 1$. Let us go back to EQ-s (21a, b). (Hereafter in the present section we discard the subscript "late"). It is convenient to consider the gauge function as depending on the scale parameter $\beta(a(t))$. Denote

$$\frac{d\beta}{da} \equiv \beta',\ \dot{\beta} = \beta' \dot{a};\ \ddot{\beta} = \beta'' (\dot{a})^2 + \beta' \ddot{a},\ b \equiv \ln \beta, b' = \frac{\beta'}{\beta}\ ;\ b'' = \frac{\beta''}{\beta} - (b')^2. \qquad (46)$$

Then in the dust dominated period ($P << \rho_m$) EQ.-s (21a, b) take the form

$$\frac{\dot{a}^2}{a^2} = \frac{8\pi}{3\beta^2}\rho_m + \frac{1}{3}\beta^2\Lambda - \frac{1}{a^2} - \left(\frac{\dot{a}}{a}\right)^2 \left[a^2 (b')^2 + 2ab'\right] \qquad (47a)$$

$$\frac{\ddot{a}}{a} = -\frac{4\pi}{3\beta^2}\rho_m + \frac{1}{3}\beta^2\Lambda - \left(\frac{\dot{a}}{a}\right)^2 \left[a^2 b'' + ab'\right] - \frac{\ddot{a}}{a}(ab') \qquad (47b)$$

In (47 a, b), $\rho_m$ stands for the matter density of luminous and Cold DM (weylons). From (47a, b) one has the following expressions for the density and pressure of DE.

$$\frac{8\pi}{3}\rho_{d.e.} = -\left(\frac{\dot{a}}{a}\right)^2 \left[a^2 (b')^2 + 2ab'\right];\qquad 4\pi P_{d.e} = \left(\frac{\dot{a}}{a}\right)^2 \left[a^2 b'' + ab'\right] + \frac{\ddot{a}}{a}(ab') \qquad (48)$$

Consider the present state with $\beta_N \equiv \beta(a_N) = 1$. It is convenient to introduce the critic density $\rho_c \equiv \frac{3}{8\pi}H^2$ ($H$ stands for the present value of the Hubble constant) as well the following density parameters: $\Omega_m = \left(\frac{\rho_m}{\rho_c}\right)_N = \frac{8\pi}{3}\frac{(\rho_m)_N}{H^2};\ \Omega_\Lambda = \frac{\Lambda}{3H^2};$

$\Omega_k = -\frac{1}{(a_N H)^2}$ and $\Omega_{d.e.} = \left(\frac{\rho_{d.e.}}{\rho_c}\right)_N = \frac{8\pi}{3}\frac{(\rho_{d.e.})_N}{H^2}$ . One can also consider the pressure parameter of DE, $\Pi_{d.e} = \left(\frac{P_{d.e.}}{\rho_c}\right)_N$ . As $a_N$ is not measurable, we introduce a flatness parameter, $\eta = a_N H$ . Large values of $\eta$ describe an almost flat universe, whereas small values – a well closed universe. We can rewrite (47a) at present as

$$\Omega_{\text{total}} \equiv \Omega_m + \Omega_\Lambda + \Omega_{d.e.} = 1 + \frac{1}{\eta^2}; \tag{49}$$

From (48) one gets

$$\Omega_\beta \equiv \Omega_{d.e.} = -\left[\left(\frac{\eta}{H}\right)^2 (b'_N)^2 + 2\frac{\eta}{H} b'_N\right] \tag{50}$$

One can make use of popular now data [27]: $H = 0.72\times10^{-28}\,\text{cm}^{-1}$; $\Lambda \approx 2.074\times10^{-58}\,\text{cm}^{-2}$**;** and the present overall $\rho_{\text{tot}} = 7.3488\times10^{-58}\,\text{cm}^{-2}$. Then $\Omega_{tot} = \frac{\rho_{\text{tot}}}{\rho_{\text{c}}} = 1.1876$. Inserting this into (49) we obtain $\frac{1}{\eta^2} = 0.1876$, i.e. $\eta \approx 2.31$. Thus, the present value of the cosmic scale parameter $a_N = \frac{\eta}{H} = 3.208\times10^{28}\,\text{cm}$.

It is believed that $\rho_{\text{tot}}$ contains: 5% ordinary matter; 23% CDM; 72% DE($\beta + \Lambda$). As $\Omega_\Lambda \equiv \frac{\Lambda}{3H^2} = 1.334\times10^{-2}$, we have $\Omega_\beta = \Omega_{DE(\beta+\Lambda)} - \Omega_\Lambda = 0.8417$. Then, from (50) we obtain two values, $(b'_N)_1 = -\frac{1}{a_N}0.6021$; $(b'_N)_2 = -\frac{1}{a_N}1.398$. Making use of EQ.(B-3) from the Appendix we have either $\delta_1 C_1 + 2\delta_2 C_2 = -0.6021$, or $\delta_1 C_1 + 2\delta_2 C_2 = -1.398$.

Let us turn to (47b). At present this may be written as

$$\left(\frac{\ddot{a}}{a}\right)_N \frac{1}{H^2}(1 + a_N b'_N) = -\frac{1}{2}\Omega_m + \Omega_\Lambda - \left[a_N{}^2 b''_N + a_N b'_N\right] \tag{51}$$

Introducing the present acceleration parameter $Q_N \equiv \left(\frac{\ddot{a}}{a}\right)_N \frac{1}{H^2}$, we can rewrite (51)

$$Q_N = -\frac{1}{2}\Omega_m + \Omega_\Lambda - (1 + Q_N)a_N b'_N - a_N{}^2 b''_N; \tag{51a}$$

As shown in the Appendix (Further to (B-3)), fixing in (51a) a fore-given $Q_N$ and combining the results with those obtained from (50) one can obtain the values of $C_1$; $C_2$; $\delta_1$; $\delta_2$ (Cf. Appendix (B-6, 7, 8))

So, choosing (arbitrary) $Q_N = 05$, one has the set

$Q_N = 0.5$: $a_N = 3.208\times10^{28}\,\text{cm}$;

$$C_1 = -10^{-2}; \delta_1 = 1.22\times10^2;\ C_2 = 10^{-2};\ \delta_2 = 3.1\times10^1 \tag{52}$$

This suitable set (52) will be adopted in the following discussion.
With the set (52) we have up to the dust period (for $a < 10^{-2} a_N$ ) $\beta_{\text{late}} \doteq 1$. Also in the early and late dust period $\beta_{\text{late}} \approx 1$. It slightly differs from 1 only in the interval $0.95 \le X \le 1.1$ (Cf. (B-9), however $\beta_{late}(X = 1) = 1$.

## 7. The dynamics in the dust dominated universe

The dust dominated period lasts from $0.01 a_N$ up to the maximum radius of the universe, $a_{\max}$. To understand the behavior of physical quantities during this period, one can go back to (47a, b) and (48). One can make use also of the energy relation (31b), that in the dust dominated period gives $\rho(a) = \frac{Const.\beta(a)}{a^3}$. At present $\beta(a_N) = 1$, so that $Const. = \rho_N a_N{}^3$, and $\rho(a) = \rho_N \beta(a) \left(\frac{a_N}{a}\right)^3$.

Taking into account that during the whole dust period $\beta_{\text{late}} \approx 1$, we rewrite (47a) as

$$\left(\frac{\dot{a}}{a}\right)^2 (1 + ab')^2 = \frac{\Lambda}{3} - \frac{1}{a^2} + \frac{8\pi}{3} \rho_N \left(\frac{a_N}{a}\right)^3 \tag{53}$$

From (53) one can obtain the maximum value of the scale parameter $a_{\max}$. Indeed, rewriting (53) $\left(\frac{\dot{a}}{a}\right)^2 (1 + ab')^2 = \frac{\Lambda}{3} - \frac{1}{a_N{}^2 X^2} + \frac{8\pi}{3} (\rho_m)_N \frac{1}{X^3}$ one verifies that $\left(\frac{\dot{a}}{a}\right)^2$ has a minimum at $X_{\max} \approx 2.66$ ; so that $a_{\max} = 2.66 a_N$.

Let us introduce the current expansion parameter $h \equiv \frac{\dot{a}}{a}$ as well the current density parameters: $\omega_m \equiv \frac{8\pi\rho_m}{3h^2} = \frac{8\pi}{3h^2} \rho_N \left(\frac{a_N}{a}\right)^3 = \Omega_m \left(\frac{H}{h}\right)^2 \left(\frac{a_N}{a}\right)^3$, $\omega_\Lambda \equiv \frac{\Lambda}{3h^2} = \Omega_\Lambda \left(\frac{H}{h}\right)^2$, that of the spatially curvature $\omega_k \equiv -\frac{1}{h^2 a^2} = -\frac{1}{\eta^2} \left(\frac{H}{h}\right)^2 \left(\frac{a_N}{a}\right)^2$ and that of DE (Cf. (48)) $\omega_{d.e.} \equiv \frac{8\pi\rho_{d.e.}}{3h^2} = -\left[a^2 (b')^2 + 2ab'\right]$ .Further, introducing the current acceleration parameter, $Q \equiv \frac{\ddot{a}}{a} \frac{1}{h^2}$ one can rewrite (47a, b) as

$$\omega_m + \omega_\Lambda + \omega_k + \omega_{d.e} = 1 \tag{53a}$$

and

$$Q(1 + ab') = -\frac{1}{2} \omega_m + \omega_\Lambda - \left[a^2 b'' + ab'\right] \tag{53b}$$

From the definitions one has $\frac{\omega_\Lambda}{\omega_m}=\frac{\Omega_\Lambda}{\Omega_m}X^3$, $\left|\frac{\omega_k}{\omega_m}\right|=\frac{1}{\eta^2\Omega_m}X$ and $\left|\frac{\omega_\Lambda}{\omega_k}\right|=\eta^2\Omega_\Lambda X^2$, so that in the late universe, $a>a_N$, the contributions of $\Lambda$ and of the spatial curvature are dominant, while in the early dust universe that of matter is dominant.

From the tables given in the Appendix and the additions after Table II one concludes that $ab'$ as well $P_{d.e.}$ both are positive and very small at $X=0.01$, then they increase and in the interval $0.995\le X\le 1.005$ they become negative; further, for $1.1\le X\le 2.66$ they take on positive values decreasing to negligible values at $X=2.66$.

The density parameter of DE $\omega_{d.e.}$ as well $\rho_{d.e.}$ both are negative and insignificant for $0.01\le X\le 0.9$. In the interval $0.995\le X\le 1.005$ they become positive and dominant and for $1.1\le X\le 2.66$ they again are negative decreasing to very small values.

The ratio $\frac{|\rho_{d.e}|}{\rho_m}$ is growing from $5\times 10^{-33}$ at $X=0.01$, up to 2.48 at $X=1$ and then it decreases to zero at $X_{\max}=2.66$.

Finally, the acceleration parameter $Q\equiv\frac{\ddot{a}}{a}\frac{1}{h^2}$ is calculated according to (53b). At $X=0.01$ one has $Q=-0.5$. $Q$ remains negative up to $X\approx 0.9$ and becomes positive in the interval $0.995\le X\le 1.005$; at $X=1$ there is $Q=0.5$. After $X=1.1$ and up to $X_{\max}=2.66$, $Q$ is negative achieving $Q=-3$ at $X_{\max}=2.66$.

## 8 An overview. Discussion

The considered model is based on the Weyl-Dirac theory that provides us with three geometrically based functions: the symmetric metric tensor $g_{\mu\nu}$, the Weyl length connection vector $w_\nu$, the Dirac gauge function $\beta$. Instead of setting in the action integral (8) for Dirac's parameter $\sigma=0$ (Cf. Dirac [19]), we make use of Rosen's approach [20], $\sigma=-2\kappa^2$, that enables to get dark matter particles.

According to our standpoint, the space-time of the universe has a chaotic Weylian microstructure but is on a large scale homogeneous and isotropic. In order to describe both, the microstructure and the global cosmic features, the Weyl vector was presented as composed of two parts, $w_\nu=\underset{\text{glob}}{w_\nu}+\underset{\text{loc}}{w_\nu}$, with $\underset{\text{glob}}{w_\lambda}=-\frac{\beta_\lambda}{\beta}$ being a global gradient vector, so that globally there are no vector fields, and with $\underset{\text{loc.}}{w_\nu}$ representing the locally restricted chaotically oriented vector fields existing in micro-cells. These local Weyl fields create bosons, named weylons, having mass $m=\frac{\kappa\hbar}{c}$ and spin 1. With $m>10\text{MeV}$ these particles form the cold dark matter bulk in the universe.

At the beginning there was no matter; the embryo universe was a homogeneous, 3-dimensional spherically symmetric geometric entity having the radius $a_0 = 1.616\times10^{-33}\text{cm}$ and "filled" with Dirac's gauge function $\beta_{\text{create}}$. Very close to the beginning occurs the process of matter creation by $\beta_{\text{create}}$ (Cf. (31a)), which continues up to the moment, when the Planck density is achieved. From this on and up to $t_1 = 1.05\times10^{-31}\text{cm}$, lasts the prematter period, characterized by the equation of state $P = -\rho$ with $\rho = \rho_{Pl}$. During this period there is an "inflation-like" growing of the radius, from $a_{\text{Pl}} = 1.2\times10^{-31}\text{cm}$ to $a_1 = 1.37\times10^{-5}\text{cm}$ and a growing of the temperature from $T_{Pl} = 2.7\times10^{-165}K$ to $T_1 = 5.24\times10^{17}K$.

The prematter period is followed by a transition period, during which prematter gradually transforms into radiation and ultrarelativistic particles. In the transition period the temperature achieves its maximum $T_{\max} = 7.4\times10^{31}K$ and decreases to $T_{\text{Rad}} = 2.43\times10^{27}K$. At these huge temperatures micro-cells of inhomogeneity appear, in which the local Weyl fields $\underset{\text{loc}}{w_\nu}$ create weylons. Following previous papers [23a, b] we presume that the ensemble of weylons constitute a cold dark matter gas. As conventionally matter (particles and radiation) has no Weylian charge it will not interact with the weylon dark matter. It is worth pointing out that after $a > A$ and up to $a_{Dust}$ one has $\beta \doteq 1$, whereas terms with $ab'$ and $a^2b''$ are extremely small (Cf. (B-10) compared to terms with matter density. Thus instead of (47a, b) we can write $\frac{\dot{a}^2}{a^2} = \frac{8\pi}{3}\rho_m + \frac{1}{3}\Lambda - \frac{1}{a^2}$ and $\frac{\ddot{a}}{a} = -4\pi\left(P_m + \frac{1}{3}\rho_m\right) + \frac{1}{3}\Lambda$ as in Einstein's GRT. This is an additional reason for setting $\beta = 1$ inside the Weyl micro-cells.

The radiation period lasts from $a_{\text{Rad}} = 100\,\text{cm}$ to $a_{\text{Dust}} \approx 3.208\times10^{26}\text{cm}$, during it the temperature of conventionally matter decreases from $T_{\text{Rad}} = 2.43\times10^{27}\,\text{K}$ to $T_{\text{Dust.conv.}} = 7.575\times10^{2}\,\text{K}$, the weylons temperature is significantly lower.

It is believed that during the early dust dominated epoch, close to $a_{\text{Dust}}$ and $T_{\text{Dust.conv.}}$, the galaxies were formed, so that plausibly the weylon DM, through its gravitational interaction with conventional matter, played an important spurring role in the galaxy formatting process.

In the dust period, the gauge function creates dark energy, which is characterized by an energy density $\rho_{d.e.}$ (Cf. (48)). The latter is insignificant in the early and late (beyond the present state) dust period, but become dominant at present with $\left(\frac{|\rho_{d.e}|}{\rho_{\text{m.}}}\right)_N = 2.48$, when it is causing cosmic acceleration.

After the present state the expanding continues up to $a_{\max} = 2.66a_N$, where one has $h^2 \approx 2.3\times10^{-59}\text{cm}^{-2}$ and a great deceleration parameter $Q \equiv \frac{\ddot{a}}{ah^2} = -3$. The latter will cause the universe to contract back to the beginning.

The model is built up with a certain Dirac gauge function Cf. (24)-(26), the latter having parameters linked to cosmological observations. It is of course possible to choose other function: so, one could take $\beta_{\text{create}} = \text{B}e^{-\gamma^2 t^2}$ or $\beta_{\text{create}} = \text{B}e^{-\gamma^2 t^2 - \delta^4 t^4}$ and obtain similar results. The function $\beta_{\text{late}}$(Cf. (26)) is also not a sole choice. In addition one could (Cf. Appendix (B-6), (B-7)) pick up a different set of parameters for $\beta_{\text{late}}$. However, the purpose of the present work was to show that on the basis of the Weyl-Dirac theory one can build up a model, where conventionally matter, DM and DE are created by geometry. This aim is achieved.

From the first glance it seems strange that an arbitrarily chosen gauge function may give rise to measurable physical phenomena. But if one is holding the opinion that the universe is a physical (observable and measurable) reality the procedure of choosing a gauge looks naturally enough. Let us recall some well-known facts. Astronomical observations led us to the cosmological principle and this directed us to the F-L-R-W line element in cosmology. Thereby a system of reference was fixed. A well known physical quantity, the mass of an insular system, has meaning only in asymptotically flat systems of reference.

Now, let us accept the existence of dark matter and the accelerated expansion at present. In the framework of Einstein's general theory of relativity there is no possibility to describe these two phenomena as stemming from geometry. But we can make use of a slightly generalized framework, the Weyl-Dirac geometry. In the latter, besides coordinate transformations (enabling the choice of a system of reference), Weyl gauge transformations are allowed. So, that we can fix a certain gauge function. In this work we have chosen the gauge on the basis of the present amount of dark matter, of the value of the deceleration parameter, and of some other plausible characteristics of the models.

**References**

[1] Steven Weinberg, *Cosmology*, Oxford University Press – 2008.
R. Durrer and R. Maartens *Gen. Rel. Grav*. 40, 301 (2008).
M. Kamionkowski, *Dark Matter and Dark Energy*, arXiv:0706.2985.
B. Ratra and M. S. Vogeley, *The Beginning and Evolution of the Universe*, arXiv:0706.1565; PASP **120**, 235 (March 2008). PASP=Publ. of the Astr. Soc. Pacific.

[2] S. Weinberg, *Gravitation and Cosmology* (Wiley, New York, 1972).

[3] A. H. Guth (1981), *Phys. Rev*. D-23, 347.

[4] A. D. Linde, *Phys. Lett.* B 108, 389 (1982); *Rep.Prog.Phys*.47. 925 (1984); arXiv:hep-th/0402051. *Phys. Scripta,* T 117, 40 (2005).
J. Garcia-Bellido and A. Linde, *Phys. Rev*.51, 429 (1995).

[5] J. A. S. Lima, A. S. M. Germano and L. R. W. Abramo, *Phys. Rev.* D 53, 4287, (1996).

[6] F. Zwicky, Helvetia Physica Acta 6, 110 (1933).

[7] S. J. Perlmutter *et al., Nature (London)* 391, 51, (1998). S. J. Perlmutter *et al., Astrophys. J.* 517, 565 (1999). A. G. Riess *et al., Astron. J.* 116, 1009 (1998). P. M. Garnavich *et al., Astrophys. J.* 509, 74, (1998).

[8] G. Lemaître, *Ann. Soc. Sinc. Brux*. Ser. I A 53, 51 (1933).

[8a] J.-L. Lehners, P. J. Steinhardt, N. Turok, arXiv:0910.0834.
Tirthabir Biswas, arXiv:0810.1315v4 (31 July 2009).

[9] M. Israelit and N. Rosen, *Astrophys. J.* 342, 627 (1989). N. Rosen and M. Israelit, *Gravitation and Modern Cosmology,* (p.151) Eds. A. Zachichi et al. (Plenum Press, New York, 1991). M. Israelit and N. Rosen, *Astrophys. Sp. Sc*.,204, 317(1993).
[10] S. P. Starkovich and F. I. Cooperstock, *Astrophys. J.* 398, 1 (1992). S. S. Bayin, F. I. Cooperstock, and V. Faraoni, *Astrophys. J.* 428, 439 (1994).
[11] H. J. Blome and W. Priester, *Astron. Astrophys.* 250, 43 (1991).
[12] E. Rebhan, *Astron. Astrophys.* 353, 1 (2000)
[13] C. S. Bohum and F. I. Cooperstock, *Phys. Rev.* A60, 4291 (1999).
G. P. Perry and F. I. Cooperstock, *Class. Qu. Grav*. 16, 1889 (1999).
[14] H. Dehnen and F. Ghaboussi, *Phys. Rev.* D 33, 2205 (1986).
F. Ghaboussi, H. Dehnen and M. Israelit, *Phys. Rev.* D 35, 1189 (1987).
H. Dehnen, F. Ghaboussi , H. Frommert, *Int. J. Theor. Phys.*, 27(5), 567 (1988).
N. M. Bezares-Roder and H. Dehnen, *Gen. Rel. Grav.* 39, 1259 (2007).
A. B. Balakin, H. Dehnen, A. E. Zayats, *Int. J. Mod. Phys.* D 17, 1255 (2008).
[15] P. S. Wesson, *Phys. Lett.* B 276, 299 (1992).
P. S. Wesson, *Astrophys. J.* 394, 19 (1992).
J. M. Overduin and P. S. Wesson, *Phys. Rept.* 283 , 303 (1997).
P. S. Wesson, *Space-Time-Matter,*(World Scientific, Singapore, 1999)
P. S. Wesson, B. Mashhoon, H. Liu, W. N. Sajko, *Phys. Lett.* B 456, 34 (1999).
S. S. Seahra and P. S. Wesson, *Gen. Rel. Grav.* 33, 1731 (2001).
H. Liu and P. S. Wesson, *Int. J. Mod.Phys.* D 10, 905 (2001).
Ponce de Leon, *Mod. Phys. Lett.* A 17, 2425 (2002).
P. Halpern and P. S. Wesson, *Brave New Universe*. (Joseph Henry Press, Washington, 2006).
[16] M. Israelit, *Found. Phys.* 35, 1725 (2005). *Found. Phys.*35, 1769 (2005).
[17] M. Israelit, *Found. Phys.* 37, 1628 (2007). *Gen. Rel. Grav*. 40, 2469 (2008).
[18] M. Israelit, *Gen. Rel. Grav.* 41, 2847 (2009).
[19] H. Weyl, *Sitzungsber. Preuss. Akad. Wiss.* , 465 (1918).
H. Weyl, *Ann. Phys.*( Leipzig) 59, 101 (1919).
H. Weyl, *Raum, Zeit, Materie,* (Springer, Berlin, 1923).
[19a] P. A. M. Dirac, *Proc. R. Soc. London* A 33, 403 (1973).
[20] N. Rosen, *Found. Phys.*12, 213 (1982).
[21] M. Israelit, *The Weyl-Dirac Theory and Our Universe.* (Nova Science, Commack, N. Y. 1999).
[21a] R. Carroll, *Remarks on Weyl Geometry and….* arXiv:0705.3921.
[22] A. L. Proca, *J. Phys. Rad*. **7,** 347 (1936)
[23a] M. Israelit and N. Rosen, *Found. Phys.* **22,** 555 (1992).
[23b] M. Israelit and N. Rosen, *Found. Phys.* **24,** 901 (1994).
[24] L. D. Landau and E. M. Lifshitz, *Statistical Physics* (Pergamon, London, 1959) p.134.
[25] S. M. Carroll, The Cosmolog. Constant, in *Living Rev. Relativity***,** (lrr-2001-1)
[26] M. Israelit, *Found. Phys*. **32,** 945 (2002).
[27] D. N. Spergel *et al.* Astrop. J Suppl. **148**, 175 (2003), ibid **170,** 377 (2007)

## Appendix

**A.** In the early universe, up to $t_A = 1.127\times10^{-31}\text{cm}$; $A = 1.544\times10^{-3}\text{cm}$, the gauge function $\beta_{\text{create}}$ is given by (25). Hereafter we discard the subscript "create".

$$\beta_{\text{create}}(t) = \frac{\mathrm{B}}{\cosh^n(\gamma t)};\quad \frac{\dot\beta}{\beta} = -n\gamma\tanh(\gamma t);\quad \frac{\ddot\beta}{\beta} = n^2\gamma^2\tanh^2(\gamma t) - \frac{n\gamma^2}{\cosh^2(\gamma t)} \tag{A-1}$$

The constants' values are: $n = 1000$, $\gamma = 1.726\text{cm}^{-1}$, $\mathrm{B} = 0.744\times10^{62}$.
At the very beginning

$$\beta_0 = \mathrm{B};\quad \left(\frac{\dot\beta}{\beta}\right)_0 = 0\ ;\quad \left(\frac{\ddot\beta}{\beta}\right)_0 = -n\gamma^2 \tag{A-2}$$

The function $\beta_{\text{create}}$ has a maximum at $t = 0$ and a flex point at $\sinh(\gamma t) = \frac{1}{\sqrt{n}}$.

For small values of $\gamma t$ ($\gamma t << 1$), one has for $\beta_{\text{create}}$ and its derivatives

$$\beta \doteq \mathrm{B};\quad \frac{\dot\beta}{\beta} \doteq -n\gamma^2 t + n\gamma^4\frac{t^3}{3};\quad \frac{\ddot\beta}{\beta} \doteq -n\gamma^2 + n\gamma^4(n+1)t^2 \tag{A-3}$$

**B.** Let us turn to, $\beta_{\text{late}}$ that acts at $t > t_A = 1.127\times10^{-31}\text{cm}$; and $a > A = 1.544\times10^{-3}\text{cm}$.

This function is given by $\beta_{\text{late}} = 1 + C_1\tanh[\delta_1(X-1)] + C_2\tanh[\delta_2(X^2-1)]$, with $X(t) = \frac{a(t)}{a_N}$ (Cf. EQ. (26)). Hereafter we will discard the subscript "late"

The derivatives are

$$\beta' \equiv \frac{d\beta}{da} = \frac{C_1\delta_1}{a_N\cosh^2[\delta_1(X-1)]} + \frac{2aC_2\delta_2}{a_N{}^2\cosh^2[\delta_2(X^2-1)]} \tag{B-1}$$

$$\beta'' = -\frac{2C_1\delta_1{}^2\tanh[\delta_1(X-1)]}{a_N{}^2\cosh^2[\delta_1(X-1)]} + \frac{2C_2\delta_2}{a_N{}^2\cosh^2[\delta_2(X^2-1)]} - \frac{8a^2C_2\delta_2{}^2\tanh[\delta_2(X^2-1)]}{a_N{}^4\cosh^2[\delta_2(X^2-1)]} \tag{B-2}$$

At present we have $\beta_{\text{late}} = 1$ and

$$b'_N = \left(\frac{\beta'}{\beta}\right)_N = \frac{1}{a_N}(C_1\delta_1 + 2C_2\delta_2) \tag{B-3}$$

$$b''_N = \frac{\beta''_N}{\beta_N} - (b'_N)^2 = \frac{2C_2\delta_2}{a_N{}^2} - \frac{1}{a_N{}^2}(C_1\delta_1 + 2C_2\delta_2)^2 \tag{B-4}$$

In order to account the parameters of $\beta_{\text{late}}$ we consider EQ. (50). Inserting on the left hand side the value $\Omega_\beta = 0.8417$ we get either $a_N b'_N = -0.6021$, or $a_N b'_N = -1.398$. These data have to be combined with EQ (51a). Substituting in (51a) $\Omega_m = 0.28\Omega_{tot} = 0.3325$, $\Omega_\Lambda = 1.334\times10^{-2}$ (Cf. the data given after EQ. (50)) and taking into account (B-3, B- 4) one obtains additional conditions on the constants

$$2C_2\delta_2 = -1.529\times10^{-1} - Q_N - (1+Q_N)a_N b'_N + (a_N b'_N)^2 \tag{B-5}$$

In addition, there is the condition $C_1 \tanh \delta_1 + C_2 \tanh \delta_2 = 0$ (Cf. EQ (45)). Considering possible (arbitrary) values of the present cosmic acceleration $Q_N$ we can write down two sets

Set I. $\delta_1 C_1 + 2\delta_2 C_2 = -0.6021$ with: (B-6)

| $Q_N$ | $C_1\delta_1$ | $C_2\delta_2$ |
|---|---|---|
| 0.1 | −1.382 | 0.39 |
| 0.25 | −1.322 | 0.36 |
| 0.5 | −1.222 | 0.31 |

Set II. $\delta_1 C_1 + 2\delta_2 C_2 = -1.398$ with: (B-7)

| $Q_N$ | $C_1\delta_1$ | $C_2\delta_2$ |
|---|---|---|
| 0.1 | −4.558 | 1.58 |
| 0.25 | −4.698 | 1.65 |
| 0.5 | −4.798 | 1.70 |

It turns out that an appropriate choice is the third row of set I, with $Q_N = 05$. Further, EQ. (45) is satisfied by

$$C_1 = -10^{-2} : \delta_1 = 1.22 \times 10^2;\; C_2 = 10^{-2};\; \delta_2 = 3.1 \times 10^1. \qquad \text{(B-8)}$$

This set of constants (B-8) is adopted. With (B-8) one accounts:

$$\beta_{\text{late}}(X < 10^{-2}) = 1;\; \beta_{\text{late}}(10^{-2} \le X \le 0.9) = 1;\; \beta_{\text{late}}(X = 0.99) = 1.003;$$
$$\beta_{\text{late}}(X = 1) = 1;\; \beta_{\text{late}}(X = 1.1) = 1.001;\; \beta_{\text{late}}(X \ge 1.2) = 1 \qquad \text{(B-9)}$$

Thus, $\beta_{\text{late}}$ is slightly different from 1 only in the interval ($0.95 \le X \le 1.1$)

Making use of (B-1, B-2) one can consider the derivatives of $\beta_{\text{late}}$. Up to the beginning of the dust period ($X \le 0.01$) one obtains for the derivatives

$$ab'(X) = 3.2 \times 10^{-28}\left(9.2X - 1.6 \times 10^{-76}\right); \qquad \text{(B-10)}$$

$$a^2 b'' = 10^{-26}\left(1.39 \times 10^{-76} + 3.03 \times 10^{-1} + 37.6X^2\right) - 10^{-55}\left(9.2X - 1.6 \times 10^{-76}\right)^2 \text{(B-11)}$$

These are extremely small. Thus, in the early universe, up to $a_{\text{Dust}}$, $\beta_{\text{late}} \doteq 1$ and the terms with $ab'$ and $a^2 b''$ in the EQ-s may be neglected.

In **TABLE I**. are listed the quantities, $F(X) \doteq a^2\beta'' - (ab')^2 + ab' \doteq a^2 b'' + ab'$, $ab' = a\dfrac{\beta'}{\beta}$ and $a^2\beta''$, which are calculated according to (B-1), (B-2) with (B-8)

**TABLE I.**

| $X = a/a_N$ | $ab'$ | $a^2\beta''$ | $F \equiv a^2\beta'' - (ab')^2 + ab'$ |
|---|---|---|---|
| 0.01 | $2.9\times10^{-31}$ | $3.0\times10^{-27}$ | $3.0\times10^{-27}$ |
| 0.1 | $5.5\times10^{-29}$ | $1.2\times10^{-26}$ | $1.2\times10^{-26}$ |
| 0.2 | $1.4\times10^{-27}$ | $2.1\times10^{-25}$ | $2.1\times10^{-25}$ |
| 0.5 | $4.0\times10^{-21}$ | $5.1\times10^{-19}$ | $5.1\times10^{-19}$ |
| 0.8 | $3.2\times10^{-10}$ | $4.0\times10^{-8}$ | $4.1\times10^{-8}$ |
| 0.9 | $1.5\times10^{-5}$ | $1.9\times10^{-3}$ | $1.9\times10^{-3}$ |
| 1.0 | $-6.0\times10^{-1}$ | $6.2\times10^{-1}$ | $-3.4\times10^{-1}$ |
| 1.1 | $1.3\times10^{-2}$ | $-8.2\times10^{-4}$ | $1.1\times10^{-2}$ |
| 1.5 | $4.3\times10^{-14}$ | $-1.5\times10^{-31}$ | $4.3\times10^{-14}$ |
| 2.0 | $1.3\times10^{-33}$ | $-2.0\times10^{-78}$ | $1.3\times10^{-33}$ |
| 2.66 | $\approx 0.0$ | $\approx 0.0$ | $\approx 0.0$ |

Table II contains the current values of the squared expansion parameter $(h)^2 \equiv \frac{\dot{a}^2}{a^2}$ as well those of the density parameter of DE $\omega_{d.e.} = \frac{8\pi\rho_{d.e.}}{3h^2} = -\left[a^2(b')^2 + 2ab'\right]$.

**TABLE II**.

| $X$ | $\omega_{d.e.}$ | $h^2\ (\mathrm{cm}^{-2})$ |
|---|---|---|
| 0.01 | $-5.1\times10^{-31}$ | $1.7\times10^{-51}$ |
| 0.10 | $-1.1\times10^{-28}$ | $1.6\times10^{-54}$ |
| 0.20 | $-2.8\times10^{-27}$ | $1.9\times10^{-55}$ |
| 0.50 | $-8.0\times10^{-21}$ | $1.3\times10^{-56}$ |
| 0.80 | $-6.4\times10^{-10}$ | $1.9\times10^{-57}$ |
| 0.90 | $-3.0\times10^{-5}$ | $1.3\times10^{-57}$ |
| 0.995 | $5.04\times10^{-1}$ | $4.9\times10^{-57}$ |
| 1.00 | $8.4\times10^{-1}$ | $5.2\times10^{-57}$ |
| 1.005 | $5.02\times10^{-1}$ | $4.5\times10^{-57}$ |
| 1.10 | $-2.6\times10^{-2}$ | $9.0\times10^{-58}$ |
| 1.50 | $-8.6\times10^{-14}$ | $1.4\times10^{-58}$ |
| 2.00 | $-2.6\times10^{-33}$ | $4.4\times10^{-59}$ |
| 2.66 | $\approx 0.0$ | $2.3\times10^{-59}$ |

In TABLE III. we present the acceleration parameter $Q \equiv \frac{\ddot{a}}{ah^2}$, the current value of DE density $\rho_{d.e.} = \frac{3}{8\pi}\omega_{d.e.}h^2$, that of conventional matter $\rho_m = (\rho_m)_N\left(\frac{a_N}{a}\right)^3$ (which includes weylon DM) as well the pressure of DE, the latter may be calculated from $P_{d.e.} = \dfrac{h^2\left(F + Q \cdot ab'\right)}{4\pi}$. The acceleration parameter $Q$ is calculated according to (53b) with $\omega_\Lambda = \dfrac{\Lambda}{3h^2}$ , and $\omega_m = \dfrac{8\pi}{3h^2}\left(\rho_m\right)_N\left(\dfrac{a_N}{a}\right)^3$.

**TABLE III**. The dynamics in the dust dominated period.

| $X \equiv \dfrac{a}{a_N}$ | $h$ $\left(\mathrm{cm}^{-1}\right)$ | $Q$ | $\rho_m$ $\left(\mathrm{cm}^{-2}\right)$ | $\rho_{d.e.}$ $\left(\mathrm{cm}^{-2}\right)$ | $P_{d.e.}$ $\left(\mathrm{cm}^{-2}\right)$ |
|---|---|---|---|---|---|
| 0.01 | $4.1\times10^{-26}$ | $-5.0\times10^{-1}$ | $2.1\times10^{-52}$ | $-1.0\times10^{-84}$ | $4.0\times10^{-79}$ |
| 0.10 | $1.3\times10^{-27}$ | $-5.3\times10^{-1}$ | $2.1\times10^{-55}$ | $-2.1\times10^{-83}$ | $1.5\times10^{-80}$ |
| 0.2 | $4.4\times10^{-28}$ | $-5.6\times10^{-1}$ | $2.6\times10^{-56}$ | $-6.4\times10^{-83}$ | $3.2\times10^{-81}$ |
| 0.5 | $1.4\times10^{-28}$ | $-3.4\times10^{-1}$ | $1.7\times10^{-57}$ | $-1.24\times10^{-77}$ | $8.0\times10^{-76}$ |
| 0.8 | $3.6\times10^{-29}$ | $-3.1\times10^{-1}$ | $4.0\times10^{-58}$ | $-1.5\times10^{-67}$ | $4.7\times10^{-66}$ |
| 0.9 | $6.0\times10^{-29}$ | $-3.0\times10^{-1}$ | $2.8\times10^{-58}$ | $-1.3\times10^{-62}$ | $5.4\times10^{-61}$ |
| 0.995 | $7\times10^{-29}$ | $+4.7\times10^{-1}$ | $2.2\times10^{-58}$ | $+3.6\times10^{-58}$ | $-2\times10^{-60}$ |
| 1.0 | $7.2\times10^{-29}$ | $+5.0\times10^{-1}$ | $2.1\times10^{-58}$ | $+5.2\times10^{-58}$ | $-2.6\times10^{-57}$ |
| 1.005 | $6.7\times10^{-29}$ | $0.1\times10^{-5}$ | $1.9\times10^{-58}$ | $2.7\times10^{-58}$ | $-1.8\times10^{-60}$ |
| 1.1 | $3.0\times10^{-29}$ | $-9.0\times10^{-1}$ | $1.5\times10^{-58}$ | $-5.0\times10^{-60}$ | $-5.0\times10^{-62}$ |
| 1.5 | $1.2\times10^{-29}$ | $-1.33$ | $6.1\times10^{-59}$ | $-1.4\times10^{-72}$ | $-1.6\times10^{-73}$ |
| 2.0 | $6.6\times10^{-30}$ | $-2.0$ | $2.5\times10^{-59}$ | $-1.3\times10^{-92}$ | $-4.5\times10^{-93}$ |
| 2.66 | $4.8\times10^{-30}$ | $-3.0$ | $2.0\times10^{-59}$ | 0 | 0 |